# Global perspectives on the energy landscapes of liquids, supercooled liquids, and glassy systems: The potential energy landscape ensemble

Chengju Wang and Richard M. Stratt

Department of Chemistry
Brown University
Providence, RI 02912

**Abstract**

In principle, all of the dynamical complexities of many-body systems are encapsulated in the potential energy landscapes on which the atoms move – an observation that suggests that the essentials of the dynamics ought to be determined by the geometry of those landscapes. But what are the principal geometric features that control the long-time dynamics? We suggest that the key lies not in the local minima and saddles of the landscape, but in a more global property of the surface: its accessible pathways. In order to make this notion more precise we introduce two ideas: (1) a switch to a new ensemble that removes the concept of potential barriers from the problem, and (2) a way of finding optimum pathways within this new ensemble. The *potential energy landscape ensemble*, which we describe in the current paper, regards the maximum accessible potential energy, rather than the temperature, as a control variable. We show here that while this approach is thermodynamically equivalent to the canonical ensemble, it not only sidesteps the idea of barriers, it allows us to be quantitative about the connectivity of a landscape. We illustrate these ideas with calculations on a simple atomic liquid and on the Kob-Andersen model of a glass-forming liquid, showing, in the process, that the landscape of the Kob-Anderson model appears to have a connectivity transition at the landscape energy associated with its mode-coupling transition. We turn to the problem of finding the most efficient pathways through potential energy landscapes in our companion paper.

# I. Introduction

A sizeable number of approaches to understanding the dynamics of complex systems – supercooled liquids, glassy materials and biomolecules – rely on the premise that the interesting dynamics is a fairly direct consequence of the geometry of the potential energy landscapes.[1-12] In a sense, this premise is indisputable for classical systems: classical trajectories are indeed determined by the potential surface on which the atoms or molecules move. However, where one takes this observation is not as clear. It has been argued, for example, that these connections are not direct enough to be useful.[13-17] Alternatively, and perhaps more usefully, one could question whether the most important geometric features, the ones that would have the most direct dynamical implications, have ever been properly identified.

It is this last point that this paper and its companion take up. In order to pursue the geodesic approach to landscape analysis that we introduce in the companion paper,[18] it is useful to study energy landscapes in the context of a new ensemble, what we shall call *the potential energy landscape ensemble*. If we refer to the complete set of configurations (coordinates) for all of the molecules in our system as **R**, and take the potential energy to be V(**R**), then the landscape ensemble is defined to be the set of all possible molecular coordinates whose potential energy lies at or below some landscape energy $E_L$, (Fig. 1)

$$\{\mathbf{R}: V(\mathbf{R}) \leq E_L\} . \qquad (1.1)$$

These subsets of the full configuration space make up an ensemble parameterized by the landscape energy in much the same way that the microcanonical and canonical ensembles select portions of phase space (sets of coordinates and momenta) depending on either the total (kinetic plus potential) energy or the temperature, respectively. As one would expect from the equivalence of the various ensembles used in statistical mechanics, all three ensembles should predict the same equilibrium behavior for any macroscopic system.[19] Still, it is worth noting the distinction from the microcanonical ensemble in particular. Besides the replacement of phase space by configuration space, the energy is an upper bound in the potential energy landscape ensemble whereas in the microcanonical ensemble all the states of the system have the same energy. These differences are sufficient to give differing forms for the probability distribution of molecular positions, at least in low-dimensional applications. The relationships between the statistical mechanics and thermodynamics of these ensembles are discussed in the Appendix.

So why should we bother considering anything but the canonical ensemble when studying potential energy landscapes? The fixed-energy constraint makes the microcanonical ensemble computationally awkward,[20] and experiments are invariably carried out at a prescribed temperature (and not with some sort of cap imposed on the potential energy). The answer lies in the differing *dynamical* perspectives suggested by the landscape ensemble. There is a common presumption in canonical treatments that when dynamics is slow it is because of the presence of high potential barriers, regions of configuration space that are allowed but occur only with low probabilities.[7-9,12,21] Within this picture it is the length of time it takes to accumulate enough energy to surmount these barriers that largely determines the rate of a process. The landscape ensemble, however, has no potential barriers. A given configuration is either allowed or forbidden. As we show in the next section, every allowed configuration in this ensemble has an equal probability, so there are no low-probability bottlenecks that set the rates of passage through the landscape.

This shift in perspective means that we can redirect our focus from the traditional emphasis on finding stationary *points* on the landscape, saddle points that separate the local minima,[3,7,8,12] to finding and characterizing the accessible *pathways* through the landscape.[18] Within the landscape ensemble perspective, pathways would be slow not because they have to climb over high barriers, but because they have to take a long and tortuous route to avoid such barriers. The landscape ensemble has a further advantage as well; it makes it natural to relate the dynamics on the potential surface to such topological features as the connectivity of a landscape.[18]

The starting point for discussing the connectivity is the observation that since there are no forbidden configurations in the canonical ensemble, we can think of every pair of canonical configurations as being connected through a continuous series of allowed configurations. In the energy landscape ensemble, by contrast, it is possible to have well-defined disconnected regions of configuration space. As one can envision from seeing Fig. 1, it is not out of the question that the "ocean" of available configuration space can be separated into isolated "lakes" by the presence of land on all sides with an elevation greater than sea level (**R** such that $V(\mathbf{R}) \geq E_L$).[22] We should note that the existence of a significant fraction of rigorously isolated configuration space is easier to prove in few-body problems than it is in a many-body context.[23] Nonetheless, the dynamical significance of such an occurrence would be clear. The set of configurations allowed by the landscape ensemble with landscape energy $E_L$ is precisely the set visited by classical trajectories with total energy $E_L$. An absence of a connecting path between two configurations is therefore an explicit sign of nonergodicity.[24]

Of course, when the landscape ensemble is used to make these kinds of connections between dynamics and potential surface geometry, the results are obtained as functions of the landscape energy $E_L$. In order to make sense of these findings we still need to be able to relate $E_L$ to the temperature, the standard experimental control parameter. Here, the equivalence of the various ensembles[19] comes to our aid. Much as one can derive a straightforward relationship between the microcanonical total energy and the canonical temperature just by appealing to the statistical thermodynamics of the microcanonical ensemble,[19] we can derive an analogous relationship between $E_L$ and the configurational temperature $T_{config}$ by appealing to the statistical thermodynamics of the energy landscape ensemble. We arrive at this relationship as well in the next section.

The remainder of this paper will be organized as follows: Section II develops the statistical mechanics and thermodynamics of the energy landscape ensemble. With the aid of the numerical methods introduced in Sec. III, we then illustrate the temperature/landscape energy connection by carrying out numerical calculations for both an ordinary atomic liquid and glass-forming binary liquid in Sec. IV. We show, in particular, that there seems to be an "ideal" non-ergodicity transition, a well-defined landscape energy (and a corresponding temperature) below which the glass-forming liquid becomes configurationally disconnected. As an interesting sidelight, we note that the statistical thermodynamics of our ensemble also suggests a remarkably facile route for calculating the configurational density of states of a liquid. We therefore also carry out illustrative calculations of this quantity in this section as well. The paper concludes in Sec. V with a few general observations.

**II. The statistical thermodynamics of the potential energy landscape ensemble**

As with any new ensemble, the first statistical mechanical quantity we need to find is $\rho(\mathbf{R})$, the probability density for each configuration $\mathbf{R}$. From the definition of the ensemble, Eq. (1.1), we know that $\rho(\mathbf{R}) = 0$ for all $\mathbf{R}$ for which $E_L > V(\mathbf{R})$, so we can find the probability distribution by first writing the entropy as an integral of the probability density over the allowed space.

$$S/k_B = - \int d\mathbf{R}\, \theta[E_L - V(\mathbf{R})] \rho(\mathbf{R}) \ln \rho(\mathbf{R}), \quad (2.1)$$

and then maximizing subject to constraint that the distribution is normalized over this same space

$$1 = \int d\mathbf{R}\, \theta[E_L - V(\mathbf{R})] \rho(\mathbf{R}). \quad (2.2)$$

Here $\theta(x) = (1, x \geq 0;\, 0, x < 0)$ is the unit step function and $k_B$ is Boltzmann's constant.

The microcanonical ensemble spans a different space, but the derivation is otherwise identical. After introducing a Lagrange multiplier $\alpha$ that enforces the normalization constraint, we require that

$$\frac{\delta \overline{S}}{\delta \rho(\mathbf{R})} = 0, \quad \overline{S}/k_B \equiv S/k_B - \alpha \int d\mathbf{R}\, \rho(\mathbf{R}), \quad (2.3)$$

giving us an answer that has the same form it does in the microcanonical case: $\rho(\mathbf{R})$ is constant over the entire allowed space. From Eqs. (2.1)-(2.3),

$$\rho(\mathbf{R}) = \theta[E_L - V(\mathbf{R})] / G(E_L). \quad (2.4)$$

$$G(E_L) = \int d\mathbf{R}\, \theta[E_L - V(\mathbf{R})]. \quad (2.5)$$

Once we know this probability density, we can then proceed to develop the statistical thermodynamics of the ensemble. The entropy, for example, comes from substituting Eq. (2.4) and (2.5) back into Eq. (2.1):

$$S = k_B \ln G(E_L),  \qquad (2.6)$$

which formally allows us to define the configurational temperature associated with each landscape energy $E_L$.

$$\frac{1}{T_{config}(E_L)} \equiv \frac{\partial S}{\partial E_L} = k_B \frac{\partial \ln G(E_L)}{\partial E_L} . \qquad (2.7)$$

Equation (2.7) is actually not an especially computationally friendly expression. One can see the difficulty by recasting the expression in terms of the normalized probability density of potential energies for a given landscape energy, $P(E; E_L)$

$$P(E; E_L) \equiv \langle \delta[E - V(\mathbf{R})] \rangle = \frac{\int d\mathbf{R}\, \delta[E - V(\mathbf{R})]\, \theta[E_L - V(\mathbf{R})]}{\int d\mathbf{R}\, \theta[E_L - V(\mathbf{R})]}$$

$$= \theta[E_L - E] \frac{g(E)}{G(E_L)} \qquad (2.8)$$

where the *density of states* $g(E)$ for the system is defined by

$$g(E) \equiv \int d\mathbf{R}\, \delta[E - V(\mathbf{R})] = \frac{\partial G(E)}{\partial E} . \qquad (2.9)$$

Equations (2.8) and (2.9) imply that Eq. (2.7) could just as well have been written as

$$\frac{1}{T_{config}(E_L)} = k_B \lim_{E \to E_L^-} P(E; E_L) . \qquad (2.10)$$

Because $P(E; E_L)$ is a normalized probability, it is reasonably straightforward to sample. However the very high dimensionality of our configuration space ($\mathbf{R}$ is 3N-dimensional for a system of N atoms) implies that it is very strongly peaked at $E = E_L$, meaning that the precise value at the peak is not a particularly easy number to come by. A much easier approach to computing the temperature actually takes advantage of this high dimensionality. In any very high dimensional space, the volume of even a thin shell near the surface of a hypersphere completely dominates the interior volume of the hypersphere.[25] Thus we expect the configurations close to the surface of our allowed volume (those for which $V(\mathbf{R}) \approx E_L$) to contribute almost all of the entropy.

$$S \approx k_B \ln g(E_L), \qquad (2.11)$$

so that the temperature is also given by the easier-to-compute alternative expression

$$\frac{1}{T_{config}(E_L)} \approx k_B \left.\frac{\partial \ln g(E)}{\partial E}\right|_{E_L}$$

$$= k_B \lim_{E \to E_L^-} \frac{\partial \ln P(E; E_L)}{\partial E}. \qquad (2.12)$$

The ease with which one can calculate this derivative has an interesting consequence: Calculating the configurational entropy itself (or equivalently, finding the density of configurational states) is going to be simply a matter of integrating the landscape-energy dependent configurational temperature. The entropy change between a landscape energy $E_L$ and a reference energy $E_r$, for example, is just[26-28]

$$\Delta S = k_B \ln \frac{g(E_L)}{g(E_r)} = \int_{E_r}^{E_L} dE \frac{1}{T_{config}(E)}. \qquad (2.13)$$

There has already been a considerable effort devoted to numerical simulation of the densities of states for complex systems, especially in the context of biomolecules,[29,30] supercooled liquids and glassy systems,[26-28,31-34] as well as in spin systems.[35,36] The recent progress in developing "flat histogram" methods[29,32,33,35,36] has been responsible for much of this success. However, the computational simplicity of this energy-landscape route to the configurational entropy seems to offer some advantages when compared to the effort required by a number of other methods. We demonstrate this landscape approach for simple and glass-forming liquids in Sec. IV.

Before we do proceed to realistic fluid systems though, it is worth seeing how the potential energy landscape formalism works with some idealized problems. Consider, first, how the landscape ensemble treats a d-dimensional harmonic oscillator, a landscape with the shape of a d-dimensional bowl,

$$V(\mathbf{x}) = \tfrac{1}{2} m\omega^2 \sum_{j=1}^{d} x_j^2 \quad , \quad \mathbf{x} = (x_1, \ldots, x_d) \; . \tag{2.14}$$

For a given landscape energy $E_L$, the configurational volume $G(E_L)$ and density of states $g(E_L)$ are related to the volume and surface area (respectively) of a d-dimensional sphere

$$G(E_L) = \frac{1}{\tfrac{d}{2}\Gamma\!\left(\tfrac{d}{2}\right)} \left(\frac{2\pi}{m\omega^2}\right)^{\tfrac{d}{2}} (E_L)^{\tfrac{d}{2}}$$

$$g(E_L) = \frac{1}{\Gamma\!\left(\tfrac{d}{2}\right)} \left(\frac{2\pi}{m\omega^2}\right)^{\tfrac{d}{2}} (E_L)^{\tfrac{d}{2}-1} \; . \tag{2.15}$$

Hence, as we noted, when d is large, the product $(\delta E)\, g$ for any shell width $\delta E$, will always be much larger than G and the probability distribution of potential energies will be very sharply peaked about $E = E_L$.

$$P(E; E_L) = \frac{d}{2}\left(\frac{1}{E_L}\right)\left(\frac{E}{E_L}\right)^{\frac{d}{2}-1} . \tag{2.16}$$

The configurational temperatures computed by our two methods, Eqs. (2.10) and (2.12), are thus related to the landscape energy by the expressions

$$E_L = \left(\frac{d}{2}\right) k_B T_{config} \quad \text{and} \quad E_L = \left(\frac{d}{2}-1\right) k_B T_{config} , \tag{2.17}$$

respectively. As expected, these two do become identical in the limit of large d. Moreover, as one might also expect from the equivalence of ensembles for macroscopic systems, a canonical ensemble calculation of the dependence of the average potential energy $\langle E \rangle$ on temperature T, gives a result of exactly the same (equipartition) form

$$\langle E \rangle = \left(\frac{d}{2}\right) k_B T . \tag{2.18}$$

For finite systems quite generally, one can find other, still different, routes to obtaining energy/temperature relationships, but they all tend to give identical results in the thermodynamic limit.[37] The one, well-known, exception occurs in the vicinity of phase transitions, a phenomenon we shall encounter when we apply these ideas to liquids.[38]

The other kind of problem we want to touch on briefly is that of a fluid of hard-spheres. Potential energy landscapes approaches typically have very little to say about systems with hard-core potentials because such systems have no minima and no stationary points in the normal sense. Methods based on looking for inherent structures and saddle points[3,7,12] are thus no longer relevant. But while it is true that realistic intermolecular potentials do not have hard-core discontinuities, it has long been recognized that much of the essential geometry of liquids is well described by such models[39] – which landscape methods should, in principle, be able to capture.[40] It is therefore worth emphasizing that hard sphere fluids do fit naturally into our particular potential energy landscape ensemble perspective. The landscape consists only of regions with energy 0 and regions with energy ∞, but if we take the landscape energy $E_L$ to be 0, Eqs. (2.1)-(2.6) all apply. More importantly, our basic notion of searching for available pathways rather than special points is still viable. Temperature, as usual, is not a meaningful thermodynamic variable with hard-spheres (although the ratio of pressure to temperature is). Nonetheless the regions of configuration space available within the ensemble do have a geometry that can be probed by methods such as those introduced in our companion paper.[18]

### III. Models and Numerical Methods

In order to illustrate the statistical thermodynamics of the ensemble, we looked at two different model systems: an ordinary single-component atomic liquid,[41] and the Kob-Andersen binary atomic mixture,[42] a well-studied[4,31,32,43,44] example of a glass-forming liquid. Both systems have truncated Lennard-Jones pair potentials: for an atom of species $\alpha$ separated from an atom of species $\beta$ by a distance r,

$$u_{\alpha\beta}(r) = u^{LJ}_{\alpha\beta}(r) - u^{trunc}_{\alpha\beta}(r) \quad , \qquad (r < r_c)$$

$$u_{\alpha\beta}(r) = 0 \quad , \qquad (r > r_c)$$

with the truncation distance chosen to be $r_c = 2.5\sigma_{\alpha\beta}$. The truncation potentials $u^{trunc}(r)$, though, are slightly different.

*single-component* $\qquad u^{trunc}_{\alpha\beta}(r) = u^{LJ}_{\alpha\beta}(r_c) + u'^{LJ}_{\alpha\beta}(r_c)(r - r_c)$

*Kob-Andersen* $\qquad u^{trunc}_{\alpha\beta}(r) = u^{LJ}_{\alpha\beta}(r_c)$ .

Here $u^{LJ}_{\alpha\beta}(r)$ is the standard Lennard-Jones potential and $u'^{LJ}_{\alpha\beta}(r)$ is its derivative.

$$u^{LJ}_{\alpha\beta}(r) = 4\varepsilon_{\alpha\beta}\left[\left(\frac{\sigma_{\alpha\beta}}{r}\right)^{12} - \left(\frac{\sigma_{\alpha\beta}}{r}\right)^6\right] \quad , \qquad u'^{LJ}_{\alpha\beta}(r) = \frac{d}{dr}u^{LJ}_{\alpha\beta}(r) \quad .$$

The single-component liquid has only a single well-depth parameter $\varepsilon$ and a single distance parameter $\sigma$ to make use of in defining a thermodynamic state. We focus on systems with $N = 256$ atoms at reduced density $\rho\sigma^3 = 1.058$, examining the behavior over a range of reduced landscape energies $E_L/\varepsilon$ (or, equivalently, reduced temperatures $k_BT/\varepsilon$). However we have also repeated these calculations with 500 and 864 atoms in order to check for finite-size effects.

The Kob-Andersen mixture has separate parameters for each interaction between particles of types A and B

$$\varepsilon_{AA} = \varepsilon, \quad \varepsilon_{BB} = 0.5\,\varepsilon, \quad \varepsilon_{AB} = 1.5\,\varepsilon$$

$$\sigma_{AA} = \sigma, \quad \sigma_{BB} = 0.88\,\sigma, \quad \sigma_{AB} = 0.8\,\sigma \ .$$

For this system we again work with N = 256 total atoms (a mixture of 205 A atoms and 51 B atoms)), and again study the behavior as a function of $E_L/\varepsilon$ (and $k_BT/\varepsilon$), but here we use a total reduced density of $\rho\sigma^3 = 1.2$.

Carrying out calculations within the potential energy landscape ensemble for either of these systems requires that we sample the many-body configuration space according to the probability distribution given by Eqs. (2.4), preferably in some fashion that does not necessitate computing the normalization integral, Eq. (2.5). The most obvious way to meet these goals is to use a (zero-temperature) Metropolis Monte Carlo random walk in which every trial configuration **R** for which $V(\mathbf{R}) \leq E_L$ is accepted and every **R** for which $V(\mathbf{R}) > E_L$ is rejected. We found that this simple procedure works quite well. Given a step size $\delta r$, the Monte Carlo moves themselves were implemented by randomly selecting an individual atom j and attempting to displace it along all three of the Cartesian directions ($\mu$ = x, y, z) by a randomly chosen distance $-\delta r < \delta r_{j\mu} < \delta r$. A complete Monte Carlo step for our N-atom systems consisted of N such single-atom attempted moves.

A few technical points: For the liquid phase of the monatomic system it was straightforward to generate the initial configurations for the Monte Carlo random walk. For each desired $E_L$, we melted an fcc lattice by moving $3 \times 10^3$ complete Monte Carlo steps with an arbitrary choice of step size. The step size was then tuned by running successive blocks of $10^3$ complete steps, computing the acceptance probability for each block, and readjusting the step size until it generated an acceptance probability of about 0.5.[45] Once the optimum step size was determined, six independent walks of $10^6$ complete Monte Carlo steps (each with its own step size) were used to generate the requisite configurations needed to compute the potential energy distribution $P(E; E_L)$ for that landscape energy.

For the binary system we can use the same procedure for most landscape energies. However for the lowest $E_L$ values, the system's equilibrated potential energy is actually less than that of an arbitrary fcc lattice. We therefore used a somewhat more elaborate procedure to generate appropriate "ground-state" configurations for our Monte Carlo walk to melt. We again started with an fcc lattice (with the A and B atoms randomly distributed), but now melted our system by means of a high-temperature ($k_BT/\varepsilon = 6.0$) $10^5$ time-step molecular dynamics run and then annealed it to $k_BT/\varepsilon = 0.1$ (rescaling the momenta to this kinetic temperature every 100 steps over a span of $10^7$ time steps, and then equilibrating for $10^7$ time steps after that). These molecular dynamics calculations were carried out using the velocity-Verlet algorithm with a time step $\delta t = 0.003\ \tau_{LJ}$ (with $\tau_{LJ} = \left(m\sigma^2/\varepsilon\right)^{1/2}$ and the atomic mass m assumed to be the same for all of our atoms).[46] The resulting configurations, in turn, were quenched to their corresponding inherent structures[2-5] by a conjugate-gradient search[47] (tolerance $3\times10^{-8}$). By starting with six independent initial momentum distributions for our initial fcc lattice we thus obtained six independent inherent structures. The same Monte Carlo procedure we outlined for the monatomic liquid was then applied to the binary system using these inherent structures in place of the initial lattice configurations

To obtain the configurational temperature from the set of configurations generated for each $E_L$, the slope of $\ln P(E; E_L)$ versus E was calculated from a linear-least-square fit spanning the range from the peak of the probability distribution to about $e^{-5}$ times the peak value. Uncertainties were estimated from the 6 different Monte Carlo random walks used for each system. Comparison calculations of the canonical ensemble average potential energies as a function of canonical temperature were performed by a standard Boltzmann Metropolis Monte Carlo random walk following the same protocol we reported for the monatomic landscape calculations.

Entropy calculations were carried out by integrating the reciprocal temperature, Eq. (2.13), using Simpson's rule, suitably generalized for irregularly spaced points.[48]

## IV. Numerical Results

The central working quantity for any equilibrium application of the potential energy landscape ensemble is the probability distribution of potential energies, $P(E; E_L)$, Eq. (2.8). As one can see from the results shown in Fig. 2 for the Kob-Andersen liquid, and as we saw for the multidimensional harmonic oscillator in Sec. II, this distribution is very strongly peaked near $E = E_L$ with a nearly linear dependence of the logarithm of the probability density on the potential energy $E$. The simple Lennard-Jones system (not shown) exhibits an almost identical behavior.

This simple behavior makes it straightforward to extract the corresponding configurational temperature using Eq. (2.12). As is evident from Figs. 3 and 4, the relationship between this configurational temperature and the landscape energy is (almost) indistinguishable from that between the canonical temperature and the canonical ensemble average potential energy. For macroscopic systems the potential-energy-landscape and canonical ensembles do indeed provide perfectly interchangeable routes to the equilibrium statistical mechanics.[19,49,50]

The one exception to this statement that one can see in these figures arises from the presence of a liquid/solid phase transition in the single-component Lennard-Jones system. In the vicinity of a first-order phase transition, different ensembles frequently offer quite different perspectives.[38] Thus in Fig. 3, we notice that the liquid and solid branches of the energy/temperature curves approach each other somewhat differently when they enter the coexistence region. In the canonical ensemble there is a single melting temperature ($k_B T/\varepsilon \approx 2.15$) associated with a range of allowed values of average potential energies (pure solid < ⟨E⟩ < pure liquid), whereas in the landscape ensemble there is a single melting landscape energy ($E_L/N\varepsilon \approx -3.45$) corresponding to a range of configurational temperatures, $1.76 < k_BT/\varepsilon < 2.33$. Moreover while each canonical average energy corresponds to a unique temperature, there are some configurational temperatures associated with two different landscape energies. We should note that there are additional effects that could, in principle, occur in this region because of the finite size (and boundary conditions) of our simulations. However an explicit comparison of the landscape results for N = 256, 500, and 864 atoms (not shown) shows that there are no noticeable finite-size effects on the scale of Fig. 3.

The Kob-Andersen system, by contrast, (Fig. 4) never crystallizes so it has no such phase-transition-induced dichotomy between ensembles. The landscape ensemble predictions perfectly match the results of previous, fairly elaborate, canonical simulations throughout the entire temperature range shown.[42,43] However, here again, the landscape ensemble offers its own perspective. Figure 4 reports results only above $T_{MCT}$, the mode-coupling temperature[51] for this system ($T_{MCT} = 0.435$, corresponding to a mode-coupling landscape energy $E_L(MCT)/N\varepsilon = -7.03$).[42] At this point canonical ensemble calculations fall prey to relaxation times too large to produce reliable results. Landscape ensemble calculations, by contrast, do not face (these same) difficulties with relaxation because they have no slow barrier crossings to hinder simulations.

Still, a more detailed look at what happens in the landscape ensemble, (Fig. 5) shows that once one starts to look below the mode-coupling landscape energy, random walks starting from different inherent structures suddenly begin to lead to different configurational temperatures. In effect, *different regions of configuration space in the landscape ensemble have become disconnected.* That the system is literally becoming localized into separate regions can be seen by focusing on the portion of the ensemble sampled from our lowest energy inherent structure (Fig. 5, insert).[4] If one were really trapped in a multidimensional harmonic well of the form of Eq. (2.14), then one would expect the excess potential energy beyond that of the inherent structure, $E_{IS}$, to obey (the large d = 3N) limit of Eq. (2.17)

$$\frac{\Delta E}{N\varepsilon} \equiv \frac{E - E_{IS}}{N\varepsilon} = \frac{3}{2}\frac{k_B T}{N\varepsilon}$$

Below the mode-coupling transition. this is precisely what one sees. The slope of the line fit through the lowest 5 points is 1.474 and the intercept occurs at - 0.005.[52]

As a final point, the fact that we have configurational temperatures as functions of our landscape energies means that we can use our ensemble as a computationally convenient way of computing the entropy (or configurational density of states) for our liquids. Entropy is not an absolute quantity in classical statistical mechanics,[53] but Eq. (2.13) allows us to compute entropies relative to some reference state. We therefore show our results for the excess configurational entropy of the Lennard-Jones liquid relative to the fcc lattice at the same density (E/Nε = -6.654) in Fig. 6 and for the Kob-Andersen liquid relative to the E/Nε = -7.000 state (our lowest state within the ergodic portion of the phase diagram) in Fig. 7.

When plotted versus configurational temperature, the single-component Lennard-Jones system's phase transition manifests itself in distinct liquid and solid branches of the excess entropy separated by a clear latent energy of fusion (TΔS). Note that when it is plotted as a function of landscape energy, the same excess entropy shows no discontinuity between the phases. The Kob-Andersen system, as expected, exhibits a simple monotonic increase in excess entropy with both landscape energy (not shown) and configurational temperature. As is clear from the figure, the latter results agree precisely with previously determined canonical ensemble literature values.[34]

## V. Concluding Remarks

In a sense, what we have called the potential energy landscape ensemble is not a particularly new idea. The ensemble itself can be thought of as a kind of hybrid of the canonical and microcanonical ensembles. As with the canonical ensemble, it spans a range of allowed energies, but as with the latter, the configurations it samples are defined thermodynamically by a single, fixed energy. Nonetheless, manipulations based on this ensemble have long been standard fixtures in textbook presentations of the microcanonical ensemble; it is commonplace, for example, to calculate the microcanonical entropy of an ideal gas by replacing microcanonical-like formulas, such as Eq. (2.11), by their landscape equivalents, Eq. (2.6).[54] However, what we believe is somewhat newer is the change in perspective that this ensemble affords when one begins to think about the connectedness of, and dynamics on, potential energy landscapes.

The most dramatic shift in how one thinks about dynamics comes from the fact that the landscape ensemble has no potential barriers. How we can understand slow dynamics when the entire notion of activated events has been removed from our vocabulary is something we take up in the companion paper.[18] The other half of the story, though, is that the landscape ensemble allows an absolute, rather than probabilistic, definition of configuration-to-configuration connectedness. That means that, unlike studies based on the canonical ensemble, it is meaningful to ask whether a given landscape has percolation transition.[22] Finding a non-percolating regime in a fluid is obviously potentially intriguing in the context of a glass-forming system because it automatically implies a loss of ergodicity[51,55,56] – one whose origin stems from the geometrical properties of the landscape alone.

To take advantage of these special features of the potential energy landscape ensemble we tried to proceed rather methodically, first deriving the configurational probability distribution for the ensemble, and then using this distribution to establish the functional equivalence of temperature and what we called the landscape energy. We showed that there was a computationally easy route to the temperature from the ensemble's probability distribution of potential energies, which, in turn, offered us a very simple way to compute the entropy of a fluid. The absence of potential barriers helped us computationally as well as conceptually here because the lack of barriers removed some of the slow relaxation problems present in the canonical ensemble.

When we applied this formalism to a simple and a glass-forming liquid, we found the kinds of relationships with the canonical ensemble one would expect at equilibrium: numerically identity between the results predicted by the ensembles except in the vicinity of phase transition. However what the landscape ensemble did do was offer us an interesting insight into the behavior of the glass-former. We found evidence that there really is a percolation transition, a sudden loss of connectedness, occurring at a critical landscape energy.

One of the striking aspects of this observation is that there does not seem to be any implicit time scale for us to see this nonergodicity. There is no obvious dependence on the cooling-rate or the measurement's frequency regime, the way there is in laboratory glass formers, just an "ideal" transition generated by the geometry of the potential energy landscape. It may also be noteworthy that this transition seems to occur at a landscape energy that corresponds precisely not only to the mode coupling transition ($k_BT/\varepsilon = 0.435$)[42] but to the "sharp change in local topography" ($k_BT/\varepsilon \approx 0.45$) previously seen by Sastry, Debenedetti, and Stillinger[4,44] in their canonical ensemble examination of the potential energy landscape of this same system. These authors' landscape results were obtained by looking at the distribution of mean-square displacements from inherent structures quenched from a finite temperature liquid, but the results seem not at all inconsistent with our landscape-ensemble percolation interpretation. It is just that the landscape ensemble may offer a more precise way of thinking about how the appearance of a potential energy landscape changes with thermodynamic conditions.

**Acknowledgements:** We thank Kenneth Schweizer, Steve Berry, and Guohua Tao for helpful discussions. This work was supported by NSF grant nos. CHE-0518169 and CHE-0131114.

**Appendix: Landscape ensembles and the microcanonical ensemble**

One way to think about the landscape ensemble and the various ensemble measures we introduced in Sec. II is to note that there are really three different, but intimately related landscape ensembles one could discuss. Just as we defined the potential energy landscape ensemble as the set of all configurations **R** for which the potential energy $V(\mathbf{R}) \leq$ some energy E, we could define a kinetic energy landscape ensemble as the set of all momenta **P** whose kinetic energy $T(\mathbf{P}) \leq E$, and a total energy landscape ensemble as the set of all phase space points (**R**, **P**) for which the Hamiltonian $H(\mathbf{R}, \mathbf{P}) = T(\mathbf{P}) + V(\mathbf{R}) \leq E$. The *volumes* associated with these three are thus

$$G(E) = \int d\mathbf{R}\ \theta[E - V(\mathbf{R})]$$

$$H(E) = \int d\mathbf{P}\ \theta[E - T(\mathbf{P})]$$

$$W(E) = \int d\mathbf{R} \int d\mathbf{P}\ \theta[E - H(\mathbf{R}, \mathbf{P})] \ . \tag{A.1}$$

But since $H = T + V$, there are direct connections between these volumes

$$W(E) = \int d\mathbf{R}\ H[E - V(\mathbf{R})] = \int d\mathbf{P}\ G[E - T(\mathbf{P})] \ . \tag{A.2}$$

In much the same way, if we define the analogous *densities of states* for these three landscape ensembles

$$g(E) = \int d\mathbf{R}\ \delta[E - V(\mathbf{R})] = dG(E)/dE$$

$$h(E) = \int d\mathbf{P}\ \delta[E - T(\mathbf{P})] = dH(E)/dE$$

$$\Omega(E) = \int d\mathbf{R} \int d\mathbf{P}\ \delta[E - H(\mathbf{R}, \mathbf{P})] = dW(E)/dE \ , \tag{A.3}$$

there are direct connections between these quantities as well

$$\Omega(E) = \int d\mathbf{R}\ h[E - V(\mathbf{R})] = \int d\mathbf{P}\ g[E - T(\mathbf{P})]$$

$$= \int_{-\infty}^{E} dE' \int d\mathbf{R} \; \delta[E' - V(\mathbf{R})] \; h(E - E')$$

$$= \int_{-\infty}^{E} dE' \; g(E') \; h(E - E') \; , \tag{A.4}$$

since $h(x) = 0$ for $x < 0$. We might also note that the volume encompassed by the total energy landscape ensemble can also be written in terms of individual densities of states:

$$W(E) = \int_{-\infty}^{E} dE' \int d\mathbf{R} \; \delta[E' - V(\mathbf{R})] \; H(E - E')$$

$$= \int_{-\infty}^{E} dE' \; g(E') \; H(E - E')$$

$$= \int_{0}^{E} dE' \; h(E') \; G(E - E') \; . \tag{A.5}$$

The normalized probability distributions of energies for the various ensembles, not surprisingly, can be written in these terms too: Given a landscape energy $E_L$, if we define

$$P_V(E; E_L) = \langle \delta[E - V(\mathbf{R})] \rangle_V$$

$$= \frac{\int d\mathbf{R} \; \delta[E - V(\mathbf{R})] \; \theta[E_L - V(\mathbf{R})]}{\int d\mathbf{R} \; \theta[E_L - V(\mathbf{R})]}$$

$$P_T(E; E_L) = \langle \delta[E - T(\mathbf{P})] \rangle_T$$

$$= \frac{\int d\mathbf{P} \; \delta[E - T(\mathbf{P})] \; \theta[E_L - T(\mathbf{P})]}{\int d\mathbf{P} \; \theta[E_L - T(\mathbf{P})]}$$

$$P_H(E; E_L) = \langle \delta[E - H(\mathbf{R}, \mathbf{P})] \rangle_H$$

$$= \frac{\int d\mathbf{R} \int d\mathbf{P}\, \delta[E - H(\mathbf{R}, \mathbf{P})]\, \theta[E_L - H(\mathbf{R}, \mathbf{P})]}{\int d\mathbf{R} \int d\mathbf{P}\, \theta[E_L - H(\mathbf{R}, \mathbf{P})]} \quad , \qquad (A.6)$$

as the probability densities for having an energy E in the potential energy, kinetic energy, and total energy landscape ensembles, respectively, then all three of these can be written in comparable ways.

$$P_V(E; E_L) = \frac{g(E)\, \theta(E_L - E)}{G(E_L)}$$

$$P_T(E; E_L) = \frac{h(E)\, \theta(E_L - E)}{H(E_L)}$$

$$P_H(E; E_L) = \frac{\Omega(E)\, \theta(E_L - E)}{W(E_L)} \quad . \qquad (A.7)$$

Although we shall not belabor it here, the configurational temperature defined in two different ways in the text (from g and from G), generalizes similarly to two different kinetic temperatures and two different phase-space temperatures. The principle result of this appendix, however, is not that there are these formal differences, but that our potential energy landscape expresssions are conceptually equivalent to, and have a computationally simple connection with, all of the microcanonical ensemble predictions for equilibrium behavior. The quantity $\Omega(E)$ is, of course, the microcanonical partition function (aside from a few constant factors). Since the kinetic energy density of states is straightforward, Eq. (A.4) gives us a direct route from the potential energy density of states to the microcanonical partition function. [A system of N atoms of equal mass m, for example, has

$$h(E) = \frac{(2\pi m)^{\frac{3N}{2}}}{\Gamma\left(\frac{3N}{2}\right)} E^{\frac{3N}{2}-1} , \qquad (A.8)$$

in close analogy with the harmonic oscillator potential energy density of states.]

**Figure Captions**

**Figure 1:** The potential energy landscape ensemble. The jagged curve is a schematic rendition of the potential energy V(**R**) of a many-body system as a function of **R**, the many-dimensional vector specifying the configuration of all of the particles. For a given landscape energy $E_L$, the ensemble consists of all of the **R**'s lying within the shaded regions.

**Figure 2:** The normalized probability density of potential energies E for the Kob-Andersen model of a glass-forming liquid. Each set of points shows the results for a different choice of landscape energy $E_L$. Numerical values in this and all succeeding figures are shown in reduced Lennard-Jones units appropriate to our N-atom systems.

**Figure 3:** Temperature (T)/potential energy (E) relationships for a simple Lennard-Jones system. The rightmost region is the liquid branch, the leftmost region is the solid branch, and the intervening shaded area is the phase coexistence region. The solid line presents the configurational temperature $T_{config}$ as a function of the landscape energy $E_L$. The squares report the relationship between the canonical temperature and the canonical-ensemble average potential energy $\langle E \rangle$. Both sets of data are shown with one-standard-deviation error bars (vertical for the landscape results and horizontal for the canonical results) but the former are too small to see outside of the coexistence region.

**Figure 4:** Temperature (T)/potential energy (E) relationships for the Kob-Andersen model of a glass-forming liquid. The stars present the configurational temperature $T_{config}$ as a function of the landscape energy $E_L$. The squares and circles report two separate literature canonical-ensemble calculations of the relationship between the canonical temperature and the canonical-ensemble average potential energy $\langle E \rangle$ for this system (KA is Ref. 42 and YK is Ref. 43). The solid line is an empirical formula derived from extrapolation of some of the canonical simulation results (Ref. 43). The insert shows an expanded view of the low-temperature/low-potential energy region of the figure.

**Figure 5:** The configurational temperature/landscape energy relationship of the Kob-Andersen model below the literature-determined mode-coupling transition (MCT). The main graph shows the landscape ensemble predictions for the reciprocal temperature versus the landscape energy (in appropriately reduced units). The four symbols represent results derived from four different inherent-structure initial conditions (with potential energies $E/N\varepsilon$ = -7.42 (s), -7.52 (t), -7.61 (l), and -7.70 (n)) with the solid line the average of all four and the error bars the one-standard-deviation results derived from these four. Above the mode-coupling temperature and landscape energy (large dashed arrows), the configurational temperatures are independent of initial conditions, but they become strongly dependent once one descends below this threshold. The YK extrapolated canonical $\langle E \rangle$ vs. T relationship (Ref. 43) is shown for reference as a short-dashed curve. The insert replots the data from the lowest inherent-structure initial conditions (n) showing the configurational temperature T as a function of the difference between the landscape energy and the initial inherent structure energy, $\Delta E$. The solid line here is a linear-least-squares fit through the lowest 5 points.

**Figure 6:** The excess entropy of the simple Lennard-Jones system as a function of the configurational temperature T (main graph) and of the landscape energy $E_L$ (insert). The excess entropy is measured here relative to an fcc crystal at this density.

**Figure 7:** The excess entropy of the Kob-Andersen system as a function of temperature. The excess entropy is measured here relative to the system at landscape energy $E_L/N\varepsilon$ = -

7.000 (corresponding to $k_BT/\varepsilon = 0.46$). The solid line presents the landscape ensemble predictions for the dependence on configurational temperature; the points are taken from a literature canonical ensemble calculation (the liquid branch reported in Ref. 34).

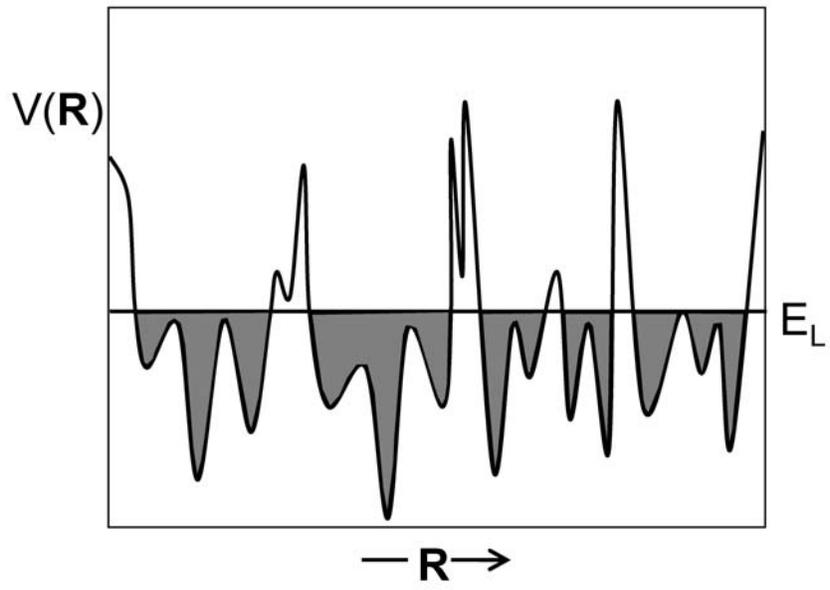

Figure 1

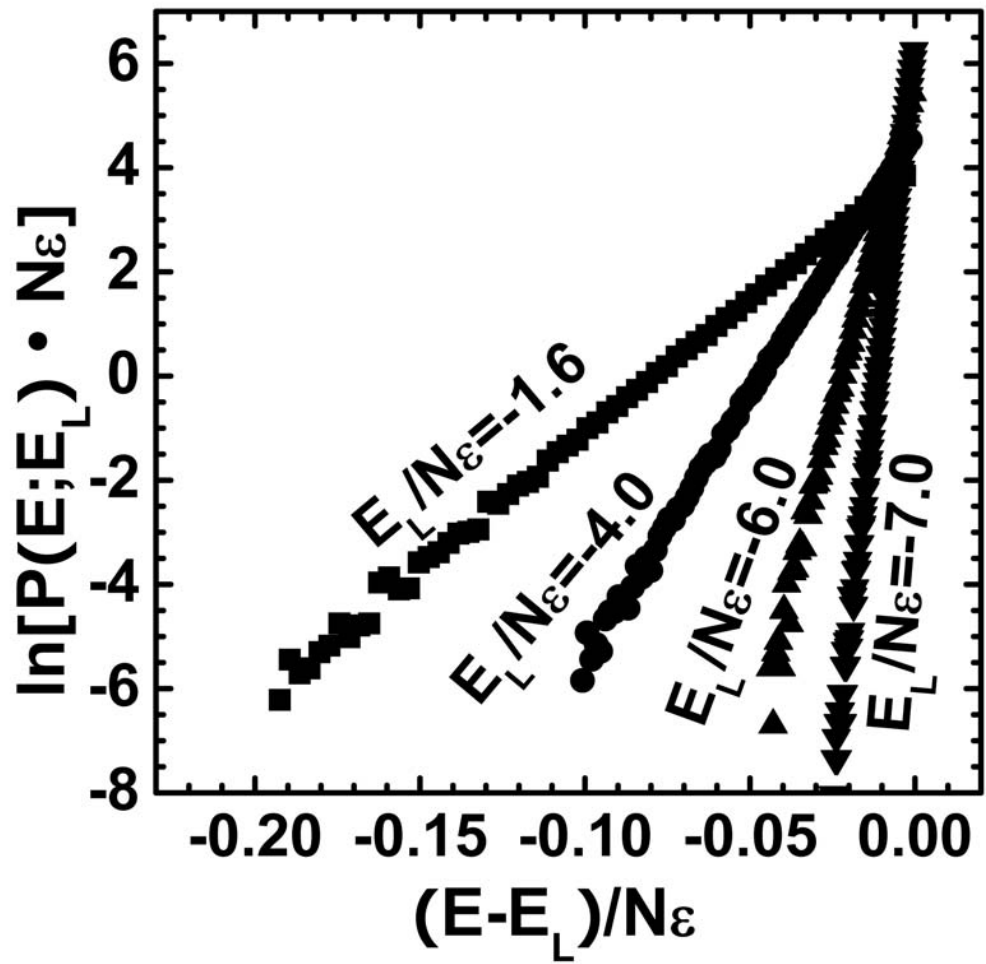

Figure 2

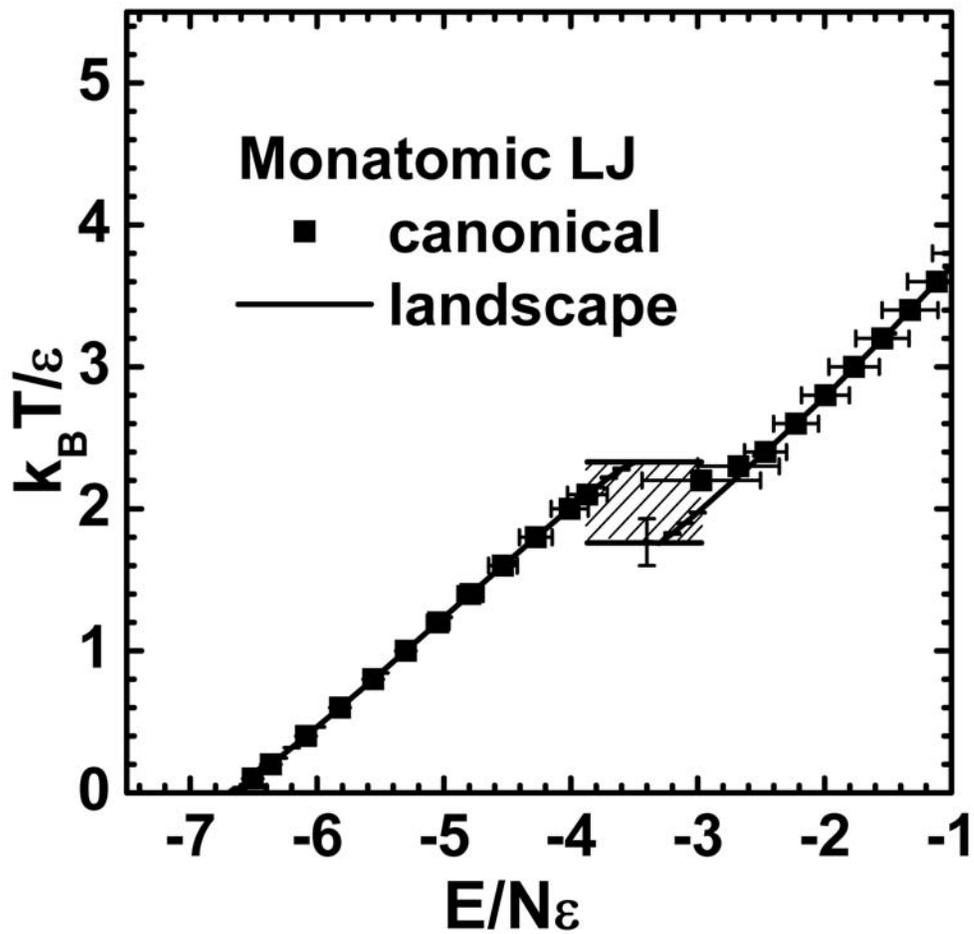

Figure 3

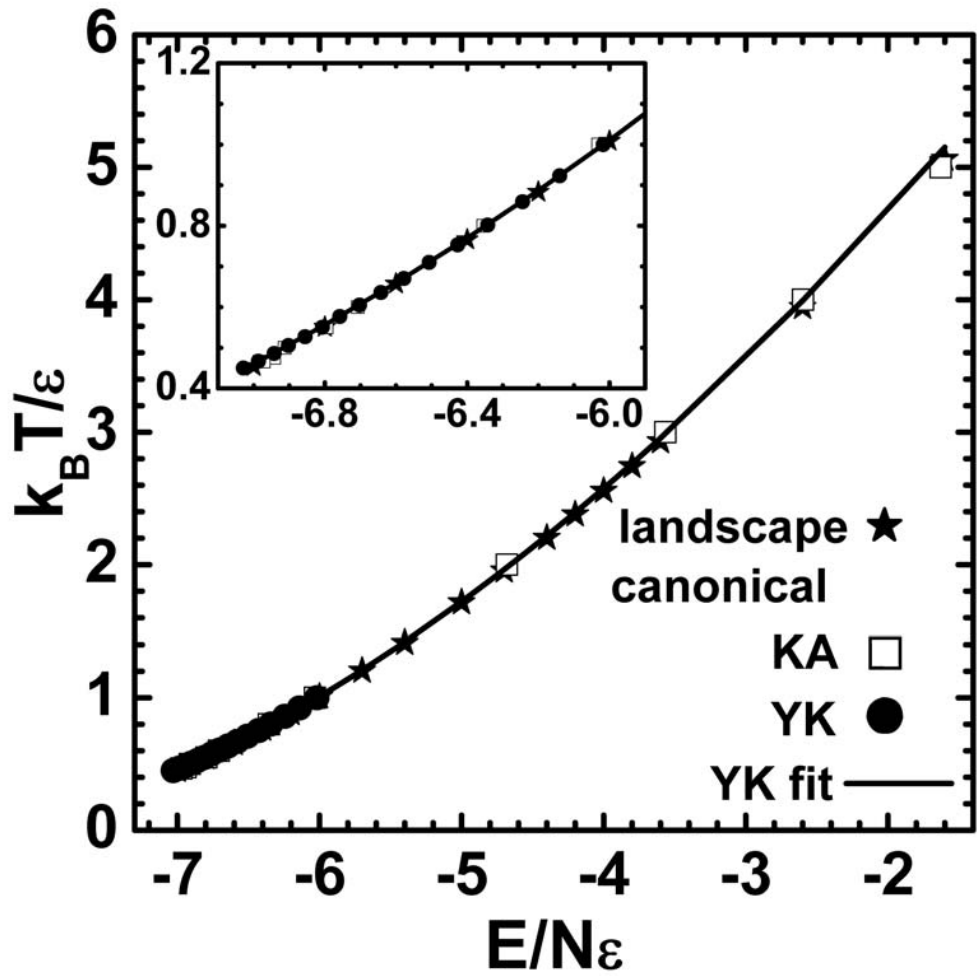

Figure 4

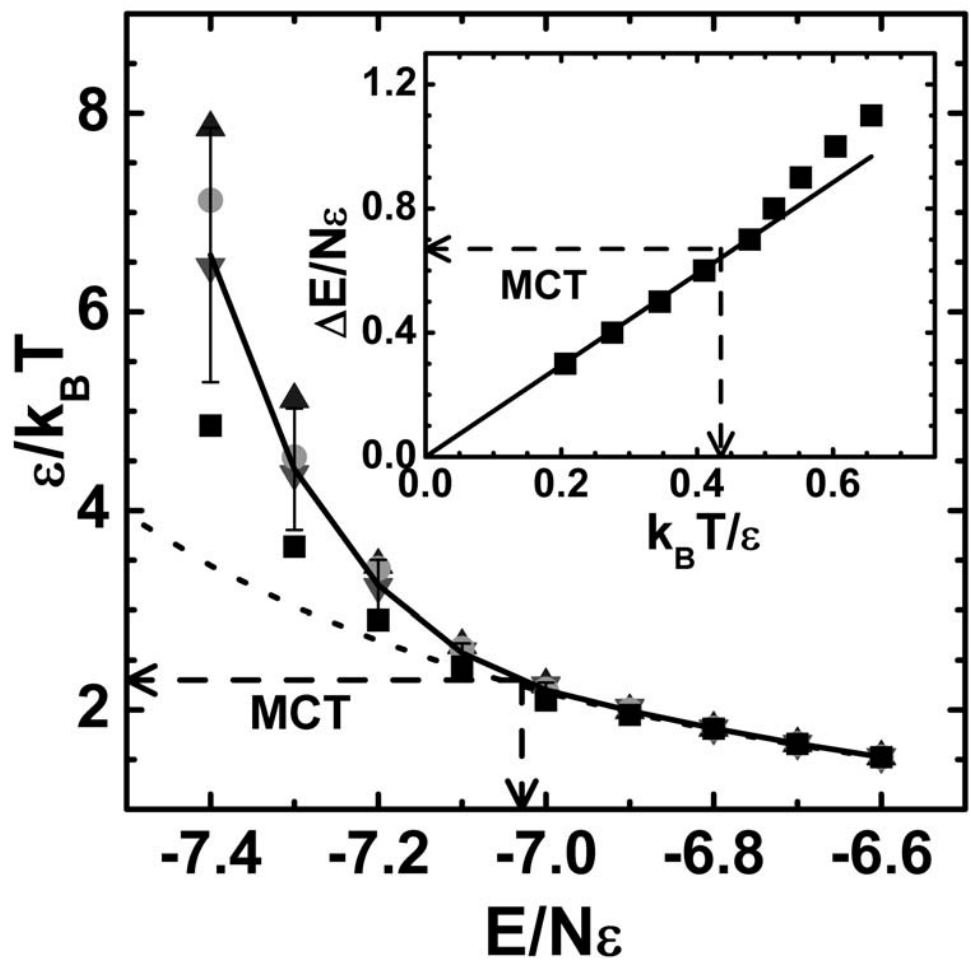

Figure 5

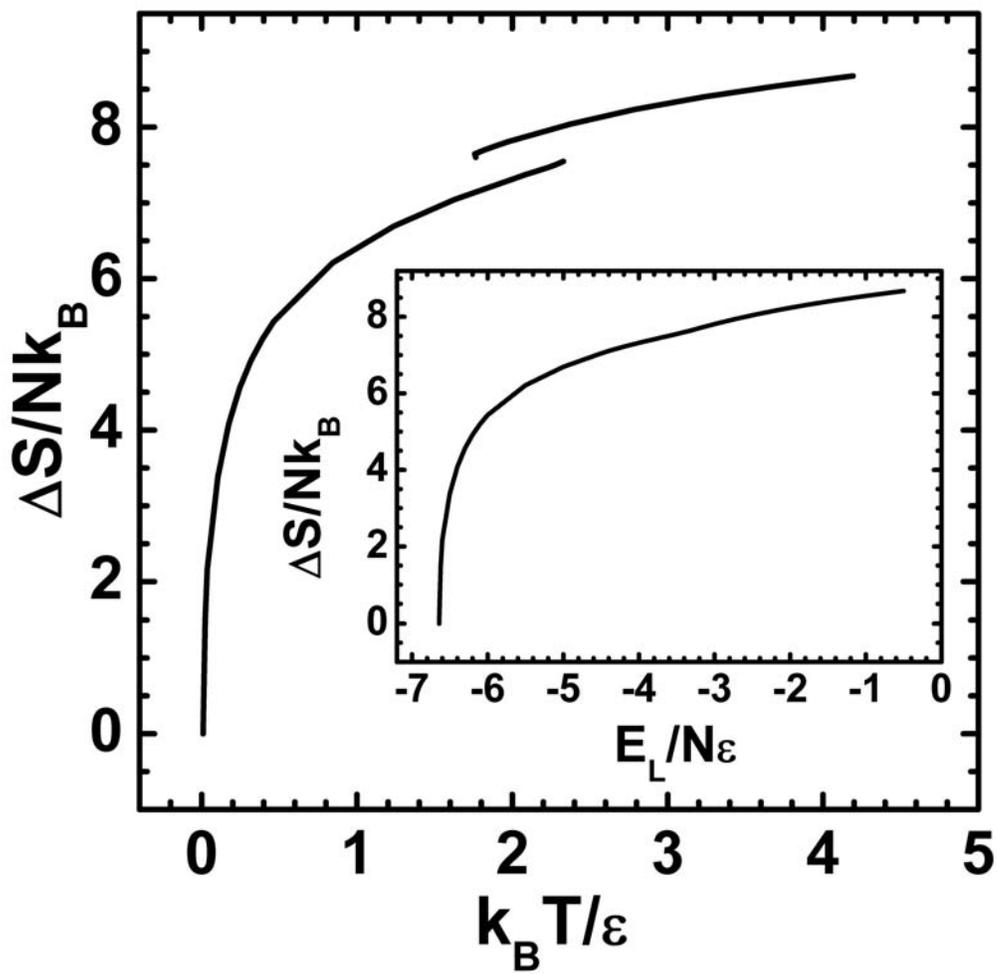

Figure 6

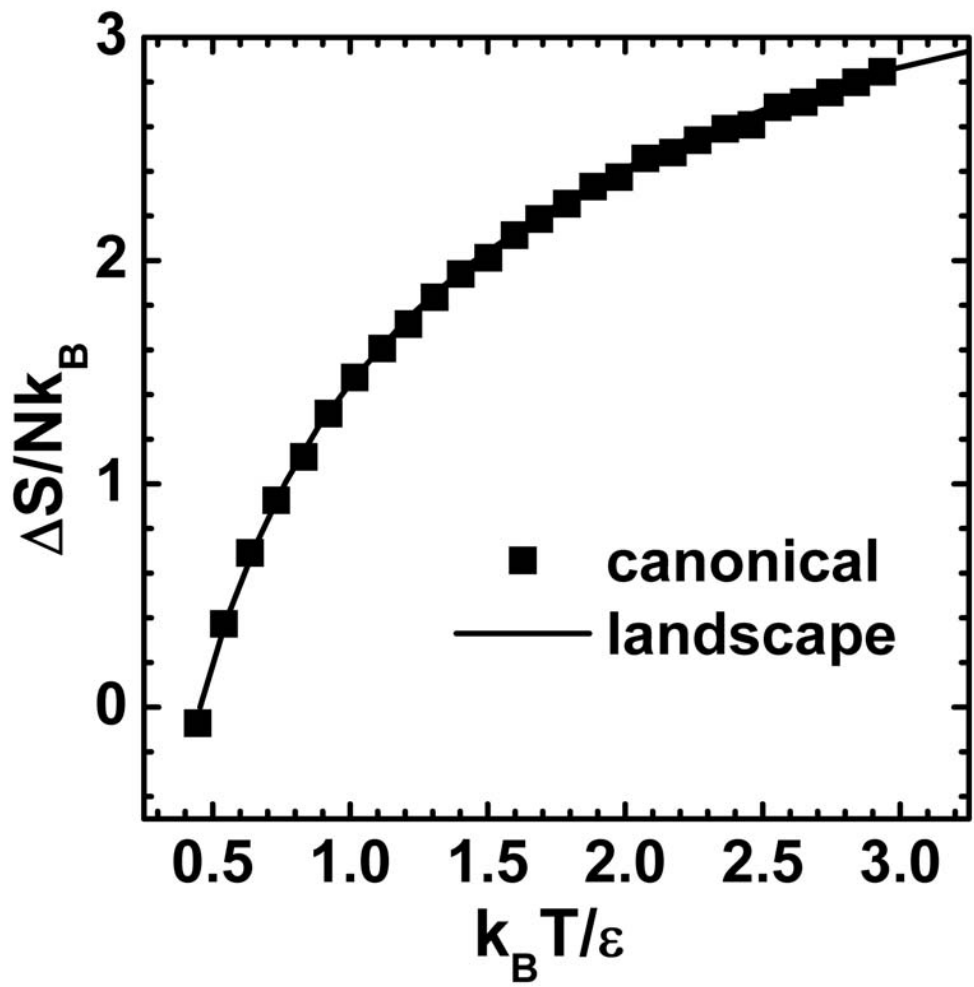

Figure 7